\documentclass[aps,14pt,prb]{revtex4}
\usepackage{amsmath,amssymb}
\usepackage{graphicx}
\usepackage{dcolumn}
\usepackage{bm}
\usepackage{color}
\usepackage{epstopdf}
\usepackage{subfigure}

\usepackage[]{natbib}

\newcommand{\mbb}{\mathbb}
\newcommand{\mc}{\mathcal}

\newcommand{\tet}{\texttt}

\begin{document}
\title{Many-Body Effects and Optical Properties of Single- and Double Layer $\alpha$-$\mathcal{T}_3$ Lattices}

\author{Andrii Iurov$^{1} \footnote{E-mail contact: aiurov@mec.cuny.edu, theorist.physics@gmail.com
}$,
Godfrey Gumbs$^{2,3}$, 
and
Danhong Huang$^{4,5}$
}

\affiliation{
$^{1}$Department of Physics and Computer Science, Medgar Evers College of City University 
of New York, Brooklyn, NY 11225, USA \\
$^{2}$Department of Physics and Astronomy, Hunter College of the City
University of New York, 695 Park Avenue, New York, New York 10065, USA\\
$^{3}$Donostia International Physics Center (DIPC),
P de Manuel Lardizabal, 4, 20018 San Sebastian, Basque Country, Spain\\
$^{4}$Air Force Research Laboratory, Space Vehicles Directorate,
Kirtland Air Force Base, New Mexico 87117, USA
$^{5}$Center for High Technology Materials, University of New Mexico,
1313 Goddard SE, Albuquerque, New Mexico, 87106, USA \\
}

\date{\today}

\begin{abstract}
An extensive analytical and numerical  investigation has been carried out to examine the role played by many-body effects on various $\alpha$-$\mathcal{T}_3$ materials under an off-resonance optical dressing field.  Additionally, we explore its dependence on the hopping parameter $\alpha$ as well as the electron-light coupling strength $\lambda_0$.  The obtained dressed states due to mutual interaction between Dirac electrons and incident light are shown to demonstrate rather different electronic and optical properties in comparison with those in the absence of incident light.  Specifically, various collective transport and optical properties of these electron dressed states are discussed in detail  and compared for both single- and double layer $\alpha$-$\mc{T}_3$ lattices.  All of these novel  properties are due to the presence of a middle flat band and the interband transitions between it and an upper conduction band.  Also, coupled plasmon dispersions for interacting double layer $\alpha$-$\mc{T}_3$ lattices are  calculated,  revealing a lower acoustic-like plasmon branch with tunable group velocity determined by both the layer separation and Fermi energy of each layer.  Finally, a many-body theory is presented within the random-phase approximation  for calculating the optical absorbance of doped multi-layered $\alpha$-$\mc{T}_3$ lattices in a linearly-polarized light field. We anticipate that the discoveries reported here could impact the design of the next-generation nano-optical and nano-plasmonic devices.  
\end{abstract}

\maketitle

\section{Introduction}

Recently, the $\alpha$-$\mc{T}_3$ lattice model\,\cite{ore121, re2, AA1} has become the object of considerable attention within the well-studied family of Dirac-cone materials. The quasiparticles of these structures have a relativistic energy spectrum \,\cite{g1, g2}.  Although $\alpha$-$\mc{T}_3$ still acquires a backbone honeycomb lattice as  their atomic structure, their properties are quite different from those of graphene due to the presence of an additional atom at the center of each hexagon, usually referred to as a ``hub.''   Physically, the hopping amplitude between this hub atom ($C$) and one of the two inequivalent ``rim'' atoms ($A$ or $B$) at the corners of each hexagon is not the same as the that between any two neighboring rim atoms. Therefore, the ratio between the hub-rim and rim-rim hopping amplitudes is chosen as a structural parameter $\alpha$ or simply taken as a geometry phase $\phi = \tan^{-1}\alpha$. In the special case of $\alpha=1$ or $\phi=\pi/4$, these two hopping amplitudes become equal to each other, corresponding to the so-called ``dice lattice''.  On the other hand, the opposite case with $\alpha = 0$ simply reduces to a completely detached hub atom plus an uncoupled hexagonal lattice like graphene.

\medskip
 
For all nonzero values of $\alpha$, the low-energy electronic states of the $\alpha$-$\mc{T}_3$ model can be determined from a $3\times 3$ pseudospin-$1$ Dirac-Weyl Hamiltonian.  This provides an extra middle flat band at the Dirac point in comparison with the Dirac-cone band structure of graphene. The flat band has proven to be stable against various types of disorder and in the presence of various boundary conditions for $\alpha$-$\mc{T}_3$-lattice  nanoribbons\,\cite{gusR1, gusR2}.    Importantly, a Berry curvature can be introduced by the structural parameter $\alpha\neq 1$ within the momentum space for conduction electrons in the $\alpha$-$\mc{T}_3$ model,  leading to an anomalous thermal-equilibrium Hall current\,\cite{ourInterplay} in addition to a thermal-equilibrium longitudinal current.

\medskip

From a materials science perspective, lattices with a flat band could be fabricated based on a variety of naturally existing materials, such as tri-layers of $SrTiO_3/SrIrO_3/SrTiO_3$ \,\cite{AA1}. The flat-band included low-energy band structures were also realized in $Hg_{1-x}Cd_{x}Te$ quantum wells with an effective value of $\alpha = 1/\sqrt{3} \backsim 0.577$  by implanting a specific doping level \,\cite{mandn}.   Additionally, artificial materials with flat energy bands were also constructed based on periodic electronic networks, spin systems under a critical magnetic-field strength\,\cite{A96},  Kagome structures\,\cite{A116}, Lieb lattices embedded within a specific substrate\,\cite{A111} and optical lattices using interferometry of laser beams.   For more inclusive references, we would refer readers to a recently published comprehensive review article about fabrication of artificial flat band materials \,\cite{ALA}.

\medskip

From the perspective of fundamental physics, all crucial properties of $\alpha$-$\mc{T}_3$, such as, single-electronic\,\cite{gus01, ourInterplay}, optical\,\cite{opt1, dey1, ourpeculiar}, magnetic\,\cite{piech, Ille1, bis1, bis2} and collective properties\,\cite{mandn, dey20, gus1}  are greatly modified by the presence of the flat band, and therefore become strongly $\alpha$ dependent.  Here, the largest modification is found in the vicinity of $\alpha = 0$\,\cite{ourNew, Moj}.     Moreover, from a technology and practical perspective, the so-called  ``{\it Floquet engineering}'', or control of the electronic dressed states through tuning light-electron interaction, both in two-dimensional lattices\,\cite{kibisAll, kiMain, fA1} and on the surfaces of three-dimensional bulk materials\,\cite{is1, is2, is3},  became a major interplayer between quantum optics and low-dimensional condensed matter physics over recent years due to advances in laser technologies.

\medskip

 We find that modification of single-electron states  depends on the polarization of the incoming radiation. Circularly polarized light leads to  opening a band gap of a few $meV$ \,\cite{kiMain, kiPRL11} and therefore a drop in the dc conductivity\,\cite{kiSREP, ourPQE16, kiMan} as well as  electron tunneling\,\cite{dip, ourTi, FA, ouranomalous}.  However, a linearly-polarized dressing field induces anisotropy\,\cite{kiSREP}  in both electron states and their dispersion, including modification of  existing anisotropy in phosphorenes\,\cite{phos1, ourJAPKi}.  Meanwhile, there might also exist a Lifshitz transition in bilayer graphene\,\cite{kiIorsh}.       

\medskip

For longitudinal plasmon excitations \,\cite{book}, there are active research investigations related to innovative two-dimensional lattices, including graphene \,\cite{wu, sds1}, gapped graphene \,\cite{pavlo},  silicene \,\cite{SilMain, ezawa}, transition metal dichalcogenides \,\cite{Sch} and dice lattices\,\cite{mandn}. The reason behind why plasmons in these materials appeal to us is their very wide frequency coverage up to the terahertz limit within the Coulomb-coupled system comprising a two-dimensional layer and a semi-infinite conductor, or the so-called {\it open systems}\,\cite{ourJAP2017}.  Specifically, a fair amount of work has been done on the finite-temperature behavior of plasmons\,\cite{sdsli, sds22, ourJAPcollective, ourThermalPRB, patel1, dadkhah, ourCond18PRB, add2, ourBook2020, hnya}, their damping\,\cite{ourT1}, and plasmon-polaritons\,\cite{ourhybridPRB}  since each of these properties could be varied independently with temperature, and then the undamped plasmon branch could extend over an even higher energy range.

\medskip
 
If the proposed modification to the single-electron states in $\alpha$-$\mc{T}_3$ lattices can be achieved by means of ``Floquet engineering'', there will be considerable  interests in the control of either optical (e.g., plasmons) or transport (e.g., spin and valley-dependent currents) properties \cite{fA2} for irradiated $\alpha$-$\mc{T}_3$ materials.  Here, the calculation of the polarization function, which describes the collective response and screening of an external potential by interacting electrons in a solid, becomes a key step.  In fact, the optical properties, including plasmon modes and optical absorption, as well as the transport properties, covering scattering rates of conduction electrons by impurities and lattice phonons, can be deduced from this calculated polarization function depending on both frequency and wave vector. Analytic  expressions for the polarization function in  various frequency and wave vector regimes are found to be crucial and the most challenging part in each of the plasmon research mentioned above. Similar studies were carried out for finite-size fullerenes\,\cite{ourbuckJAP, ourbuckWJP}.       

\medskip

The outline of the rest of this paper is as follows.  In  Sec.\,\ref{sec2}, we briefly review the single-electron states of $\alpha$-$\mc{T}_3$ lattices and their energy dispersion with an emphasis on the  new states corresponding to the middle flat band at the Dirac point.  This also serves to establish the notation we used subsequently.    Following this, in Sec.\,\ref{sec3}, we discuss various properties of electron dressed states for the most general elliptically polarized incident light, as employed in Ref.\,\cite{ourpeculiar},   including the limiting case with circularly-polarized and off-resonance dressing field in detail.  The theory and calculation of the polarization function, plasmon dispersion and wave function overlap are presented in Sec.\,\ref{sec4} for a pseudospin-$1$ Hamiltonian. We also present and discuss the dependence on the structural parameter $\alpha$ in our calculated optical conductivity, coupled plasmon modes in double layer $\alpha$-$\mc{T}_3$ lattices as well as the optical absorbance in Secs.\ \ref{sec5}$-$\ref{sec7}, correspondingly.  Finally, conclusions of this paper are drawn in Sec.\,\ref{sec8} along with some discussions and remarks.      

\section{$\alpha$-$\mathcal{T}_3$ and Dice Lattice Models: Basic Electronic Properties}
\label{sec2}

The energy dispersions of electronic states of a pseudospin-$1$ $\alpha$-$\mc{T}_3$ lattice next to the two inequivalent valley points $K$ and $K'$ are determined by 
a $(3\times 3)$ low-energy Hamiltonian matrix which explicitly depends on the structural parameter $\alpha$, or $\phi = \tan^{-1} \alpha$, and is given by\,\cite{dey1, ourInterplay}  

\begin{equation}
\label{mainH}
\mbb{H}_0 (\mbox{\boldmath$k$} \, \vert \, \tau, \phi) = \hbar v_F \left[
\begin{array}{ccc}
0 & k^\tau_- \,   \cos \phi & 0 \\
k^\tau_+ \, \cos \phi & 0 & \,  k^\tau_- \, \sin \phi   \\
0 & k^\tau_+ \, \sin \phi  & 0
\end{array}
\right]\ ,
\end{equation}
where $k^\tau_\pm = \tau k_x \pm i k_y$ with $\tau=\pm$ labeling two valleys and $v_F$ denoting the Fermi velocity. 
\medskip

Two solutions for Eq.\,\eqref{mainH} are found as

\begin{equation}
\varepsilon^{\gamma=\pm 1}_{\tau, \, \phi}(\mbox{\boldmath$k$}) =  \gamma \hbar v_F k
\end{equation}
with $\gamma = - 1$ ($\gamma = + 1$) for the valance (conduction) band, while the rest one is 

\begin{equation}
\varepsilon^{\gamma=0}_{\tau, \, \phi}(\mbox{\boldmath$k$}) = 0 \ ,
\end{equation}
which turns into a flat band. Here, all three bands are independent of phase $\phi$ or $\alpha$.  
\medskip 

Moreover, the corresponding wave functions for the valence and conduction bands are obtained as

\begin{equation}
\label{Eig1}
\Psi^{\gamma=\pm 1}_{\tau, \, \phi}(\mbox{\boldmath$k$})  = \frac{1}{\sqrt{2}} \left[
\begin{array}{c}
\tau \cos \phi \,\, \tet{e}^{- i \tau \theta_{ \bf k}}  \\
\gamma \\
\tau \sin \phi \,\, \tet{e}^{+ i \tau \theta_{ \bf k}} 
\end{array}
\right]\ ,
\end{equation}
where $\theta_k = \arctan (k_y/k_x)$, while that for the flat band becomes 

\begin{equation}
\label{Eig2}
\Psi^{\gamma=0}_{\tau, \, \phi}(\mbox{\boldmath$k$}) = \left[
\begin{array}{c}
\sin \phi \,\, \tet{e}^{- i \tau \theta_{\bf k}}  \\
0 \\
- \cos \phi \,\, \tet{e}^{+ i \tau \theta_{\bf k}} 
\end{array}
\right]\ . 
\end{equation}
\medskip

As a special case, for a dice lattice with $\phi = \pi/4$, the Hamiltonian matrix in Eq.\,\eqref{mainH} reduces to\,\cite{mandn} 

\begin{equation}
\label{DmainH}
\mbb{H}_\tau^{D}(\mbox{\boldmath$k$}) = \frac{\hbar v_F}{\sqrt{2}} \left[
\begin{array}{ccc}
0 & k^\tau_- & 0 \\
k^\tau_+ & 0 & \,  k^\tau_- \\
0 & k^\tau_+ & 0
\end{array}
\right]\ ,
\end{equation}
and the wave functions in Eqs.\,\eqref{Eig1} and \eqref{Eig2} become 

\begin{equation}
\label{DEig1}
\Psi^{\gamma=\pm 1}_{\tau, \, D}(\mbox{\boldmath$k$})  = \frac{1}{2} \left[
\begin{array}{c}
\tau \, \tet{e}^{- i \tau \theta_{ \bf k}}  \\
\sqrt{2} \, \gamma \\
\tau \, \tet{e}^{+ i \tau \theta_{ \bf k}} 
\end{array}
\right]\ ,
\end{equation}
and 

\begin{equation}
\label{DEig2}
\Psi^{\gamma=0}_{\tau, \, D}(\mbox{\boldmath$k$}) = \frac{1}{\sqrt{2}} \, \left[
\begin{array}{c}
\tet{e}^{- i \tau \theta_{\bf k}}  \\
0 \\
- \tet{e}^{+ i \tau \theta_{\bf k}} 
\end{array}
\right]\ . 
\end{equation}
Different from Eqs.\,\eqref{DEig1} and \eqref{DEig2}, the components of wave functions in Eqs.\,\eqref{Eig1} and \eqref{Eig2} for general $\alpha$-$\mc{T}_3$ lattice clearly depend on phase $\phi$. Therefore, we know the resulting overlap of wave functions as well as other quantum-mechanical observables  will also rely on $\phi$. 

\section{Electron Dressed States: Circularly-Polarized Light}
\label{sec3}

In this section, we focus on the derivation and discussing properties of the so-called {\it electron-photon dressed states}, which appear  due to strong interaction of an Dirac electron in $\alpha$-$\mc{T}_3$ lattice with an external off-resonant dressing field having a frequency much higher than the characteristic energies of our system.\,\cite{kiSREP, kibisAll} Specifically, we consider a light field in the form

\begin{equation}
\label{ellA}
\mbox{\boldmath$A$}^{(E)}(t) = \left\{
\begin{array}{c}
A^{(E)}_x (t) \\
A^{(E)}_y (t) 
\end{array}
\right\} = \frac{\mc{E}_0}{\omega} \left\{
\begin{array}{c}
\cos (\omega t) \\
\beta \, \sin (\omega t )
\end{array}
\right\} \, , 
\end{equation}
where $\beta = \sin \Theta_e$ represents the ratio between field amplitudes along two axes of a polarization ellipse. In particular, the circularly polarized light corresponds to a limiting case with $\beta = 1$. 
\medskip 

In the presence of incident light, the new Hamiltonian can be obtained by a canonical substitution for the wave vector \mbox{\boldmath$k$} through

\begin{equation}
k_{x,y} \Longrightarrow k_{x,y} - \frac{e}{\hbar} \, A^{\,(E)}_{x,y}(t) \, . 
\end{equation}
Consequently, Hamiltonian in Eq.\,\eqref{mainH} will acquire an additional {\it interaction} term $\mbb{H}_I^{(e)}(t\vert\,\tau,\phi)$ and becomes

\begin{equation}
\label{NewH}
\mbb{H}_0 (\mbox{\boldmath$k$}\, \vert \, \tau, \phi) \Longrightarrow \mc{H}(\mbox{\boldmath$k$}, t \, \vert \, \tau, \phi) = 
\mbb{H}_0 (\mbox{\boldmath$k$}\, \vert \, \tau, \phi) + \mbb{H}_I^{(e)}(t\vert\,\tau,\phi)\ .
\end{equation}
Here, the second term of the new Hamiltonian in Eq.\,\eqref{NewH} is calculated explicitly as\footnote{ 
There was a typographical error in Ref.\,\cite{ourpeculiar} where this Hamiltonian was first derived.}

\begin{eqnarray}
\nonumber
&&\mbb{H}_I^{(e)}(t\vert\,\tau,\phi) = - \tau c_0 \, \sqrt{\cos^2 (\omega t) +\left[\beta \, \sin (\omega t) \right]^2}\\
&&\times 
\left[
\begin{array}{ccc}
0\ \  & \tet{exp}\left[-i \tau \, \Omega^{\,\beta} (t) \right] \, \cos \phi & 0 \\
\tet{exp}\left[+ i \tau \, \Omega^{\,\beta} (t) \right] \, \cos \phi  & 0 &
\tet{exp}\left[-i \tau \, \Omega^{\,\beta} (t) \right] \, \sin \phi \\
0\ \  & \tet{exp}\left[i \tau \, \Omega^{\,\beta} (t) \right] \, \sin \phi & 0 
\end{array}
\right]\, ,\ \ \ \ \ \ 
\label{AH}
\end{eqnarray}
where $c_0=e{\cal E}_0v_F/\omega$
quantify the strength of electron-light interaction (with the unit of energy). 
Moreover, we have introduced  in Eq.\,\eqref{AH} the  notation $\Omega^{\,\beta} (t) = \tan^{-1} \left\{\beta\tan(\omega t) \right\}$.
Equation\ \eqref{AH} can be greatly simplified for a dice lattice with $\phi = \pi/4$. For all our future calculations, 
we would like to limit the consideration with circularly-polarized light, i.e., taking $\beta = 1$.
\medskip 

The energy eigenvalue for the flat band remains to be zero, while the dispersions of the valence and 
conduction bands are modified as   

\begin{equation}
\label{zero}
\varepsilon_\gamma(k,\lambda_0) = \gamma\sqrt{\left(\lambda_0 c_0 / 2 \right)^2 + \left[ \hbar v_F k \, (1 -
\lambda_0^2 / 4) \right]^2 \,}\ ,
\end{equation}
where $\lambda_0=c_0/\hbar\omega$ represents the dimensionless electron-light coupling parameter, and $\gamma=\pm 1$ for electrons ($+$) and holes ($-$). 
From Eq.\,\eqref{zero}, we know $\varepsilon_\gamma(k,\lambda_0)$ is independent of $\phi$ although its corresponding wave function could.
For a finite wave vector $\mbox{\boldmath$k$} = \{k_x, k_y\}$ and a small electron-photon coupling constant $\lambda_0 \ll 1$, 
Eq.\,\eqref{zero} can be further approximated as 

\begin{eqnarray}
\nonumber
\varepsilon_\gamma(k,\lambda_0) / \gamma&=&\hbar v_F k - \left\{ \frac{1}{4} \hbar v_F k - \frac{1}{8} \, \frac{c_0^2}{\hbar v_F k} \right\}
\, \lambda_0^2\\
\label{approx}
&+& c_0^2 \, \left\{ \frac{1}{32} \frac{1}{\hbar v_F k} - \frac{1}{128} \, \frac{c_0^2}{(\hbar v_F k)^3} \right\} \, \lambda_0^4 + \cdots \ .
\end{eqnarray}
\medskip

For dice lattice with $\phi=\pi/4$, its wave function is obtained from diagonalizing the eigenvalue equation of Hamiltonian in Eq.\,\eqref{NewH} as

\begin{equation}
\label{Twf1}
\Psi^\tau_\gamma (\mbox{\boldmath{$k$}},\lambda_0) = \frac{1}{\sqrt{\mc{N}^\tau_{\gamma}}} \, \left[
\begin{array}{c}
\tau \, \mc{A}^\tau_{1,\gamma} \, \tet{e}^{ - i \tau \theta_{\bf k}} \\
\mc{A}^\tau_{2,\gamma} \\
\tau \,(\hbar v_F k)^2 \, \tet{e}^{ + i \tau \theta_{\bf k}} 
\end{array}
\right]\ ,
\end{equation}
where

\begin{eqnarray}
\nonumber
&&  \mc{A}^\tau_{1,\gamma}(k,\lambda_0) = (\hbar v_F k)^2 + 2 \left(
\delta_{\lambda}^2 -  \gamma\,\tau\delta_{\lambda}\sqrt{
(\hbar v_F k)^2 + \delta_{\lambda}^2}\right)\ , \\
\nonumber 
&&  \mc{A}^\tau_{2,\gamma}(k,\lambda_0) = \sqrt{2} \, \gamma \, (\hbar v_F k) \, \left(
\sqrt{
(\hbar v_F k)^2 + \delta_{\lambda}^2
} - \gamma \, \tau\delta_{\lambda}
\right) \ , \\
\label{amplit} 
&& \mc{N}^\tau_{\gamma}(k,\lambda_0 \ll 1) \backsimeq 
4 \,(\hbar v_F k)^4 - 4 \gamma\,\tau \, c_0 \lambda_0 \, (\hbar v_F k)^3 + 3 \left[
c_0 \lambda_0 \,(\hbar v_F k) \right]^2 +\cdots\ \ .\ \ \ \ \ \ 
\end{eqnarray}
Here, the new parameter $\delta_\lambda = 2 \lambda_0 c_0 / (4 - \lambda_0^2)$ is related but not exactly equal to the dressed-state energy gap 
$E_{G}\equiv 2\Delta_0=\lambda_0 c_0$. 
\medskip

\begin{figure} 
\centering
\includegraphics[width=0.6\textwidth]{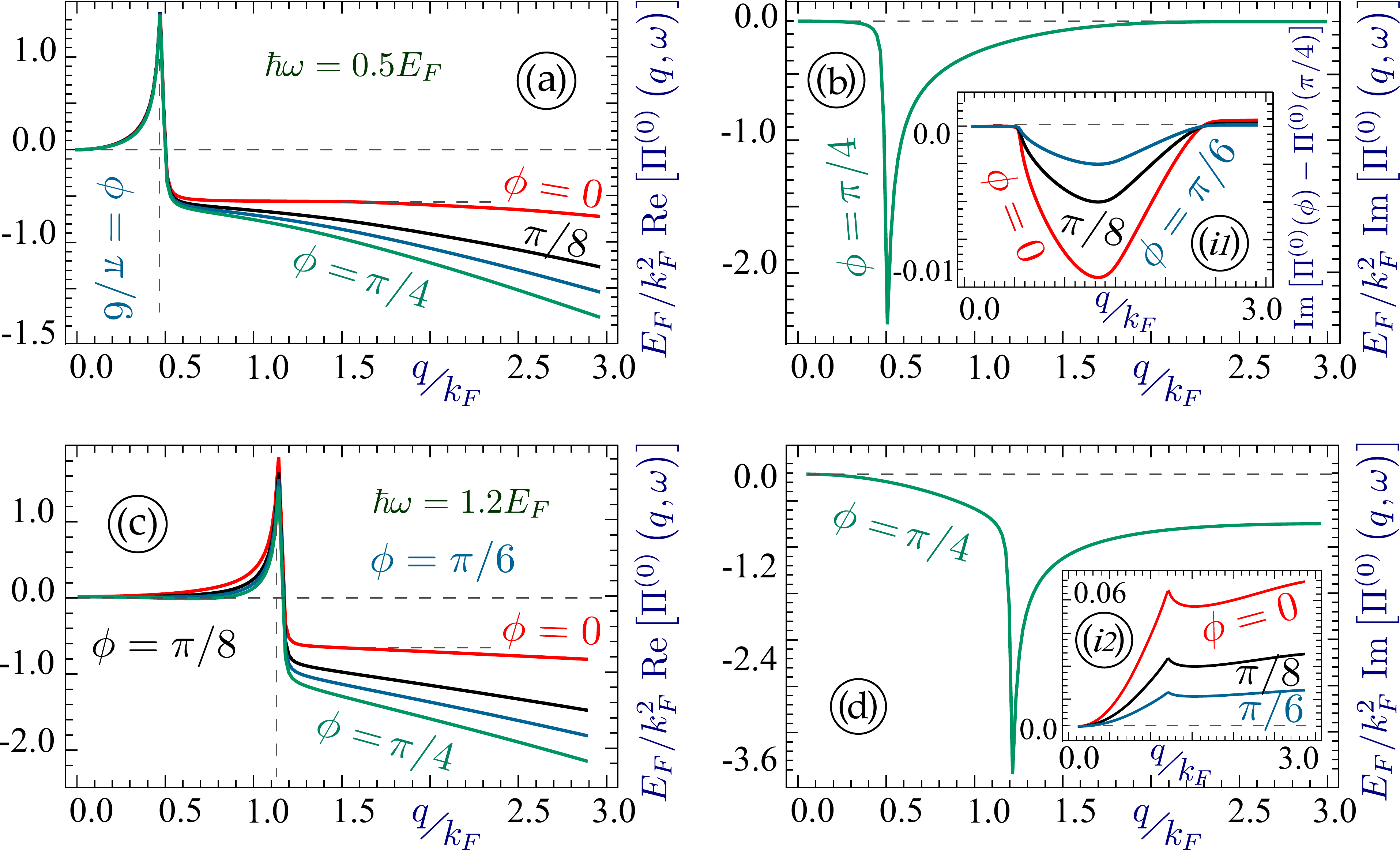}
\caption{(Color online) Polarization function $\Pi^{(0)}(q, \omega \, \vert \, \phi)$ in the units of $k_F^2/E_F$, calculated from Eq.\,\eqref{pi0at3}, for $\lambda_0=0$ and various
$\alpha$-$\mc{T}_3$ lattices as a function of scaled wave vector $q/k_F$ (Fermi wave vector $k_F$). Two left panels describe the real part of $\Pi^{(0)}(q, \omega \, \vert
\, \phi)$, while the right ones are for its imaginary part. The upper panels $(a)$ and $(b)$ correspond to a fixed frequency $\omega = 0.5\,
E_F/\hbar$, while the lower panels $(c)$ and $(d)$ show the results for $\omega = 1.2\,E_F/\hbar$. For all plots, different color curves (red, black, blue and green) 
are associated with various values of $\phi$ ($0,\,\pi/8,\,\pi/6$ and $\pi/4$), respectively. 
Insets $(i1)$ and $(i2)$ demonstrate $q$ dependence of the difference
between the imaginary part of polarization functions of $\alpha$-$\mc{T}_3$ and dice lattice, i.e., $\text{Im}\left[\Pi^{(0)}(q, \omega \, \vert \, \phi)\right]$ - $\text{Im}\left[\Pi^{(0)}(q, 
\omega \, \vert \, \pi/4)\right]$.}
\label{FIG:1}
\end{figure}
For a non-resonant field, $\lambda_0$ is expected to be small, and then, the amplitudes $\mc{A}^\tau_{1,\gamma}(k,\lambda_0)$ and 
$\mc{A}^\tau_{2,\gamma}(k,\lambda_0)$ in Eq.\,\eqref{amplit} could be expanded as 

\begin{eqnarray}
\nonumber 
&&  \mc{A}^\tau_{1,\gamma}(k,\lambda_0) \approx (\hbar v_F k)^2 - \gamma \tau \, c_0 \, (\hbar v_F k) \,\, \lambda_0 + 
\frac{1}{2} \, c_0^2 \,\, \lambda_0^2 +\cdots\ \ , \\
&&  \mc{A}^\tau_{2,\gamma}(k,\lambda_0) = \sqrt{2} \, \gamma \, (\hbar v_F k) \, \left(
\sqrt{
	(\hbar v_F k)^2 + \delta_{\lambda}^2
} - \gamma \, \tau\delta_{\lambda}
\right) \ .
\end{eqnarray}

Therefore, the dressed-state wave function in Eq.\,\eqref{Twf1} for the flat band can be approximated as

\begin{equation}
\label{psi0}
\Psi_0^\tau(\mbox{\boldmath{$k$}},\lambda_0) = \frac{1}{\sqrt{\mc{N}^\tau_{\gamma=0}}} \, \left[
\begin{array}{c}
\hbar v_F k \, \tet{e}^{ - i \tau \theta_{\bf k}} \\
2 \sqrt{2} \, c_0\lambda_0/(4 - \lambda_0^2) \\
- \hbar v_F k \, \tet{e}^{ + i \tau \theta_{\bf k}} 
\end{array}
\right] \ ,
\end{equation}
where

\begin{equation}
\mc{N}^\tau_{\gamma=0}(k,\lambda_0 \ll 1) \backsimeq 2  
(\hbar v_F k)^2 + \frac{1}{2} \, \left(\lambda_0c_0 \right)^2    + ... \ . 
\end{equation}
Compared with Eq.\,\eqref{DEig2}, we find that once the irradiation is applied, the first and last components of wave function are no longer equal to each other (except for 
a phase difference) while the middle one becomes finite $\backsim \lambda_0^2$, as excepted from a finite energy gap\,\cite{ourpeculiar,kiSREP}.

\section{Plasmons and Collective Properties}
\label{sec4}

Most of the many-body electronic properties of a solid can be obtained from the calculated polarization function $\Pi^{(0)}
(q,\omega \, \vert \, \phi,\lambda_0)$ at low temperatures with Fermi energy $E_F=\hbar v_F k_F$ with Fermi wave vector $k_F$, 
which represent a collective response of a system of many Coulomb-interacting electrons to a external potential. For an $\alpha$-
$\mc{T}_3$ lattice, $\Pi^{(0)} (q,\omega \, \vert \, \phi,\lambda_0)$ takes the form   

\begin{eqnarray}
\nonumber
\Pi^{(0)} (q,\omega \, \vert \, \phi,\lambda_0) && = \,\, \frac{1}{\pi^2} \int d^2\mbox{\boldmath{$k$}} \sum\limits_{\gamma,\gamma' = 0, \pm 1}  \,
\mbb{O}^\tau_{\gamma, \gamma'}(\mbox{\boldmath{$k$}},\mbox{\boldmath{$k$}} + \mbox{\boldmath{$q$}}\,\vert\,\phi,\lambda_0)\\
\label{pi0at3} 
&& \times \,\, \frac{\Theta[E_F-\varepsilon_{\gamma} (k,\lambda_0)] - \Theta[E_F-\varepsilon_{\gamma'}(\vert\mbox{\boldmath{$k$}} + \mbox{\boldmath{$q$}}\vert,\lambda_0)]}
{\hbar(\omega+i0^+)+ \varepsilon_{\gamma}(k,\lambda_0) - \varepsilon_{\gamma'}(\vert\mbox{\boldmath{$k$}} + \mbox{\boldmath{$q$}}\vert,\lambda_0)} \ ,
\end{eqnarray}
where $\varepsilon_{\gamma} (k,\lambda_0)$ are the dressed-state energies given by Eq.\,\eqref{zero}, $\gamma = \pm 1$  corresponds to an
electron or a hole state, and the Heaviside function $\Theta(x)$ is the limiting form of Fermi-Dirac distribution at zero temperature.
The overlap function $\mbb{O}^\tau_{\gamma, \gamma'}(\mbox{\boldmath{$k$}},\mbox{\boldmath{$k$}} + \mbox{\boldmath{$q$}}\,\vert\,\phi,\lambda_0)$ introduced in Eq.\,\eqref{pi0at3}
is simply a dot product of two wave functions $\Psi^{\gamma}_{\tau,\phi}(\mbox{\boldmath{$k$}},\lambda_0)$ and $\Psi^{\gamma'}_{\tau,\phi}(\mbox{\boldmath{$k$}}',\lambda_0)$ 
with the wave vectors $\mbox{\boldmath$k$}$ and $\mbox{\boldmath$k$}' =\mbox{\boldmath$k$}+\mbox{\boldmath$q$}$ 
as well as the band indices $\gamma$ and $\gamma'$ (also referred to the initial and scattered states)\,\cite{ourNew}, i.e.,

\begin{eqnarray}
\label{SO}
&&  \mbb{O}^\tau_{\gamma, \gamma'}(\mbox{\boldmath{$k$}},\mbox{\boldmath{$k$}} + \mbox{\boldmath{$q$}}\,\vert\,\phi,\lambda_0) = 
\Big| 
\left\langle \,
\Psi^{\gamma}_{\tau,\phi}(\mbox{\boldmath{$k$}},\lambda_0)
\,\Big|\Psi^{\gamma'}_{\tau,\phi}(\mbox{\boldmath{$k$}} + \mbox{\boldmath{$q$}},\lambda_0) \, 
\, \right\rangle 
\Big|^2 \ , 
\end{eqnarray}
where $k'= \sqrt{k^2 + q^2 + 2 k q \cos\beta_{{\bf k},{ \bf k}'}}$ is the magnitude of vector $\mbox{\boldmath$k$}'$, 
and $\beta_{{ \bf k},{ \bf k}'} = \theta_{{\bf k}'} - \theta_{\bf k}$ is the angle between $\mbox{\boldmath$k$}$ and $\mbox{\boldmath$k$}'$.
\medskip

\begin{figure} 
\centering
\includegraphics[width=0.75\textwidth]{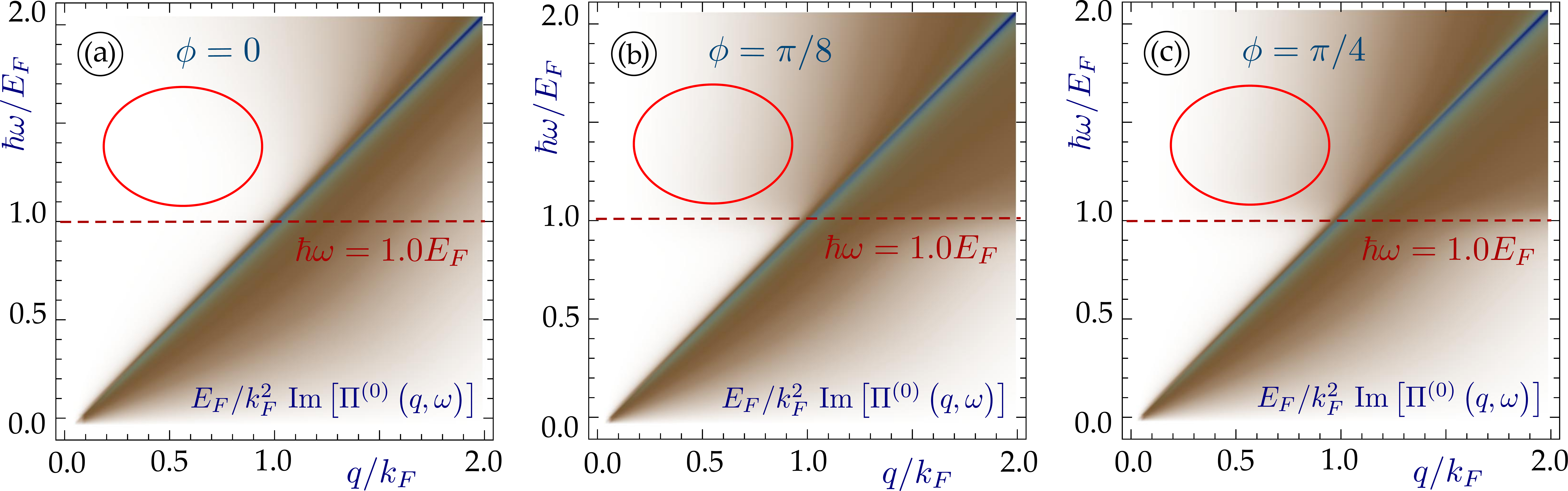}
\caption{(Color online) Density plots of $\text{Im}\left[ 
\Pi^{(0)}(q, \omega \, \vert \, \phi)\right] \neq 0$ in the units of $k_F^2/E_F$ for $\lambda_0=0$ and various $\alpha$-$\mc{T}_3$ lattices. Each
panel corresponds to a different value of structural parameter $\alpha$ or phase $\phi$: $\phi = 0$ (graphene) in $(a)$, $\phi
=\pi/8$ in $(b)$, and $\phi = \pi/4$ (dice lattice) in $(c)$.}
\label{FIG:2}
\end{figure}
The real part of $\Pi^{(0)}(q, \omega \, \vert\, \phi,\lambda_0)$ takes leading role in shaping the plasmon branches determined {\it as the zeros of the dielectric function} 
$\epsilon(q, \omega \, \vert\, \phi,\lambda_0)$ defined as     

\begin{equation}
\label{eps}
\epsilon(q, \omega \, \vert\, \phi,\lambda_0) = 1 - \textsc{v}(q) \, \Pi^{(0)}(q, \omega \, \vert\, \phi,\lambda_0) \ , 
\end{equation}
where $\textsc{v}(q) = e^2/2\epsilon_0\epsilon_rq$ is the two-dimensional Coulomb potential with $\epsilon_r$ as the host-material dielectric constant. 
Alternatively, the regions of non-zero imaginary part of $\Pi^{(0)}(q, \omega \, \vert\, \phi,\lambda_0)$
specify the particle-hole excitation regions where plasmon modes are Landau damped and break into single-particle excitations. 
Therefore, we will concentrate only on the regions of the long-living plasmon modes with $\text{Im}\left[\Pi^{(0)}(q, \omega \, \vert\, \phi,\lambda_0) \right] = 0$. 
In the random-phase approximation\,\cite{ourNew} (RPA), the dielectric function in Eq.\,\eqref{eps} can be employed to calculate Coulomb-renormalized polarization function 
$\Pi_{RPA}(q, \omega \, \vert\, \phi,\lambda_0)$, i.e.,

\begin{equation}
\Pi_{RPA}(q, \omega \, \vert\, \phi,\lambda_0)=\frac{\Pi^{(0)}(q, \omega \, \vert\, \phi,\lambda_0)}{\epsilon(q, \omega \, \vert\, \phi,\lambda_0)}\ ,
\label{RPA}
\end{equation}
where the plasmon dispersion, determined by $\text{Re}\left[\epsilon(q, \omega \, \vert\, \phi,\lambda_0)\right]=0$,
can be mapped out by the peak in the density plot of $\Pi_{RPA}(q,\omega \, \vert \, \phi,\lambda_0)$ as a function of $\omega$ for each given $q$.
\medskip 

As a first step, we focus on the polarization function $\Pi^{(0)}(q, \omega \, \vert\, \phi)$ for $\lambda_0=0$, 
plasmons and the energy loss function for non-irradiated $\alpha$-$\mc{T}_3$ lattices with different values of $\alpha$. 
Our numerical results for $\Pi^{(0)}(q, \omega \, \vert\, \phi)$ are presented in Fig.\,\ref{FIG:1}, where 
both real and imaginary parts of $\Pi^{(0)}(q, \omega \, \vert\, \phi)$ display a significant enhancement with increasing structural parameter $\alpha$. 
As already mentioned, the most drastic variation of all optical properties occurs in the vicinity of $\alpha = 0$, 
while for the other limit of $\phi = \pi/4$ these changes are really small. 
As found from Fig.\,\ref{FIG:1}, the peaks of both real and imaginary parts of $\Pi^{(0)}(q, \omega \, \vert\, \phi)$, corresponding to the pole (denominator approaching zero) 
of Eq.\,\eqref{pi0at3}, remain independent of $\alpha$, which is expected for energy dispersions in Eq.\,\eqref{zero}.    
The difference between the imaginary parts of $\Pi^{(0)}(q, \omega \, \vert\, \phi)$ with two $\alpha$ values could be either negative or positive, 
as demonstrated in two insets of Fig.\,\ref{FIG:1} for two peak values: $\omega=0.5\,E_F/\hbar$ in $(i1)$ and $\omega=1.2\,E_F/\hbar$ in $(i2)$.
\medskip

Next, we look into the imaginary part of $\Pi^{(0)}(q, \omega \, \vert\, \phi)$ in Eq.\,\eqref{pi0at3} in order to identify the regions of the stable plasmons free from Landau damping $\text{Im}\left[\Pi^{(0)}(q, \omega \, \vert\, \phi) \right]\neq 0$, corresponding to the light (uncolored) areas within the $\omega$-$q$ plane in Fig.\,\ref{FIG:2}. 
We first notice that plasmons are not damped in the triangular region above the main diagonal $\omega = v_F q$ but below the Fermi energy $\omega = E_F/\hbar$ for all values of $\alpha$. 
The area above the Fermi level acquires substantial plasmon damping for all non-zero $\alpha$ in contrast to that of graphene with $\alpha=0$, 
where the particle-hole mode boundary is known as\,\cite{wu} $\omega = v_F(2k_F-q)$.    
\medskip

\begin{figure} 
\centering
\includegraphics[width=0.75\textwidth]{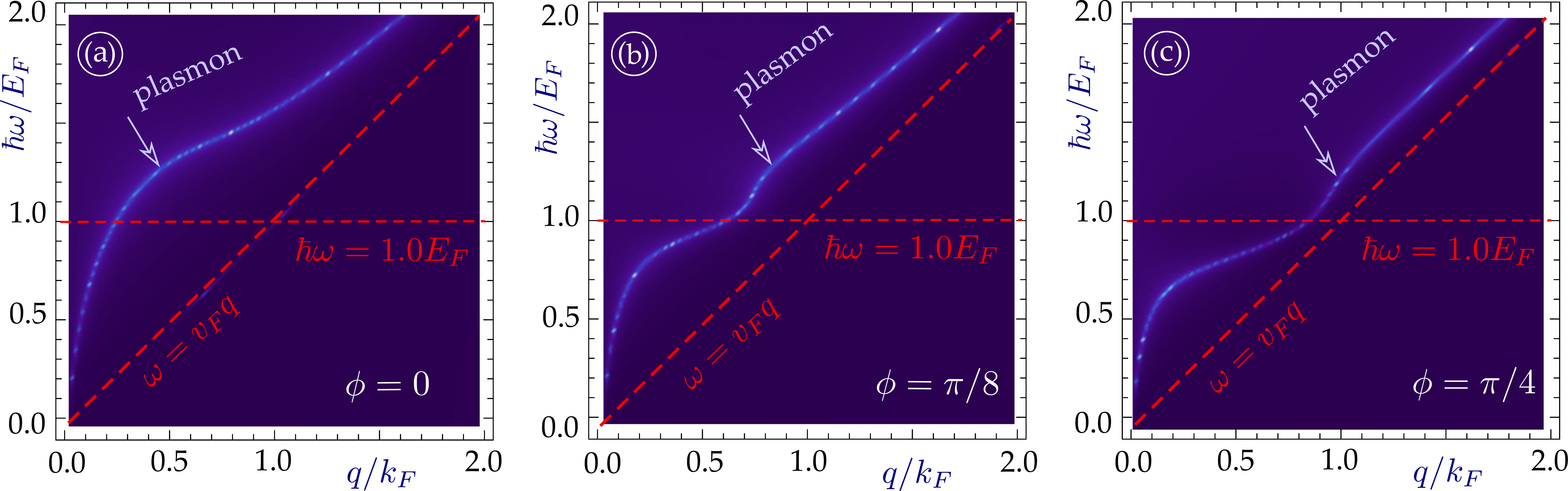}
\caption{(Color online) Density plots of $\Pi_{RPA}(q, \omega \, \vert \, \phi)$ presented in Eq.\,\eqref{RPA} in the units of $k_F^2/E_F$
for $\lambda_0=0$ and various $\alpha$-$\mc{T}_3$ lattices. Here, the plasmon dispersion (blue curve) is mapped out by the peak
in $\left|\Pi_{RPA}(q,\omega \, \vert \, \phi)\right|$ as a function of $\omega$ for each given $q$. Three panels correspond to a different values of structural parameter $\alpha$ or phase
$\phi$: $\phi = 0$ (graphene) in $(a)$, $\phi = \pi/8$ in $(b)$, and $\phi = \pi/4$ (dice lattice) in $(c)$.}
\label{FIG:3}
\end{figure}
After the calculation of unscreened polarization function $\Pi^{(0)}(q, \omega \, \vert \, \phi)$ for $\lambda_0=0$, we are able to get the screened one 
$\Pi_{RPA}(q, \omega \, \vert \, \phi)$ from Eq.\,\eqref{RPA} based on the random-phase approximation, which is presented in Fig.\,\ref{FIG:3}. 
The shape of plasmon branches for all $\alpha$-$\mc{T}_3$ lattices with $\alpha \neq 0$ in $(b)$ and $(c)$,
i.e., having two separate ``kinks'' with a much smaller separation from the main diagonal $\omega = v_F q$,
is very different from that of graphene plasmon demonstrated in $(a)$ for $\alpha=0$. 
Importantly, this separation becomes the smallest at the Fermi level with its specific values depending on  the dielectric constant $\epsilon_r$ of the host material and structural parameter $\alpha$. 
Especially, such a feature leads to a pinching of the plasmon branch in $(c)$ for a dice lattice\,\cite{mandn}. 
For all other intermediate values of $\alpha$ ($0 < \alpha < 1$), no pinching shows up as seen in $(b)$. Instead, the plasmon branch reaches the closest separation from the diagonal line $\omega=v_Fq$
as $\omega = E_F/\hbar$.   
On the other hand, in contrast to graphene, the plasmon branch for all nonzero finite $\alpha$ is free from Landau damping only limited to a small triangular region below the Fermi level 
while above the main diagonal simultaneously. This conclusion holds true even in the presence of circularly-polarized irradiation, as can be verified from Fig.\,\ref{FIG:5} below. 
\medskip

\begin{figure} 
\centering
\includegraphics[width=0.75\textwidth]{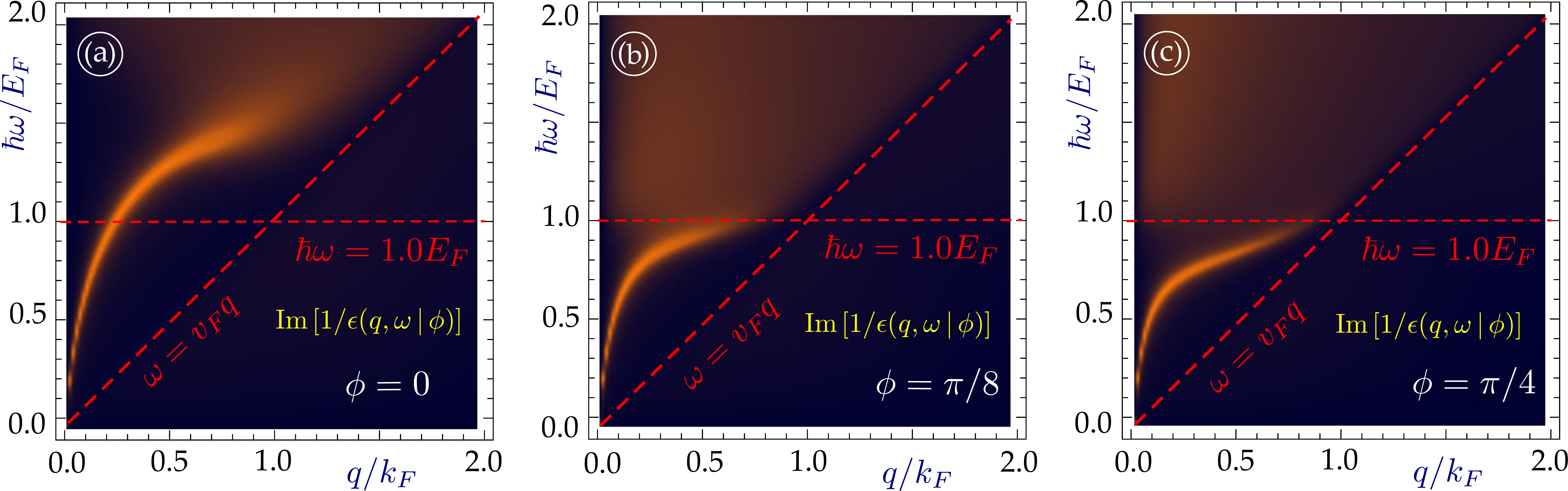}
\caption{(Color online) 
Density plots of energy loss function $\text{Im}\left[ 1/\epsilon(q, \omega \, \vert \, \phi) \right]$ given by Eq.\,\eqref{eloss} 
for $\lambda_0=0$ and various $\alpha$-$\mc{T}_3$  lattices. Here, each panel
corresponds to a different value of parameter $\alpha$ and phase $\phi$: $\phi = 0$ (graphene) in $(a)$, $\phi = \pi/8$ in $(b)$, 
and $\phi = \pi/4$ (dice lattice) in $(c)$.}
\label{FIG:4}
\end{figure}

In order to get a glimpse on kinetic-energy loss of incident charged particles, we would also calculate the electron energy-loss spectrum which 
is determined mainly by the imaginary part of the inverse dielectric function and given by 

\begin{eqnarray}
\nonumber
&& \mc{L}(q,\omega \, \vert \, \phi,\lambda_0) = \text{Im}\left[ \frac{1}{\epsilon(q,\omega \, \vert \, \phi,\lambda_0)} \right] \\
\label{eloss}
&& = \frac{ \textsc{v}(q) \, \text{Im}\left[ \Pi^{(0)}(q,\omega \, \vert \, \phi,\lambda_0) \right]}{ \left\{ 1 - \textsc{v}(q) \, \text{Re}\left[ 
\Pi^{(0)}(q,\omega \, \vert \, \phi,\lambda_0) \right] \, \right\}^2 + \left\{ \textsc{v}(q) \, \text{Im}\left[ 
\Pi^{(0)}(q,\omega \, \vert \, \phi,\lambda_0) \right] \, \right\}^2 }\ .  
\end{eqnarray}  
The plots for numerical results on the loss function $\mc{L}(q,\omega \, \vert \, \phi)$ in Eq.\,\eqref{eloss} with $\lambda_0=0$ are presented in Fig.\,\ref{FIG:4}. 
The single bright orange curve features an undamped plasmon branch, while the extended region of brownish color is associated with various energy losses due to Landau damping to plasmon. 
From Figs.\,\ref{FIG:4}$(b)$ and \ref{FIG:4}$(c)$ for energy-loss function $\mc{L}(q,\omega \, \vert \, \phi)$, 
we once again demonstrate that all plasmon branches with nonzero $\alpha$ values will be strongly damped above the Fermi level. 
The physical nature of energy loss\,\cite{Antonios1} has been revealed by Eq.\,\eqref{eloss} with its magnitude determined by the imaginary part of the polarization function 
while its dispersion by the first term of the denominator in Eq.\,\eqref{eloss}.       
\medskip

\begin{figure} 
\centering
\includegraphics[width=0.5\textwidth]{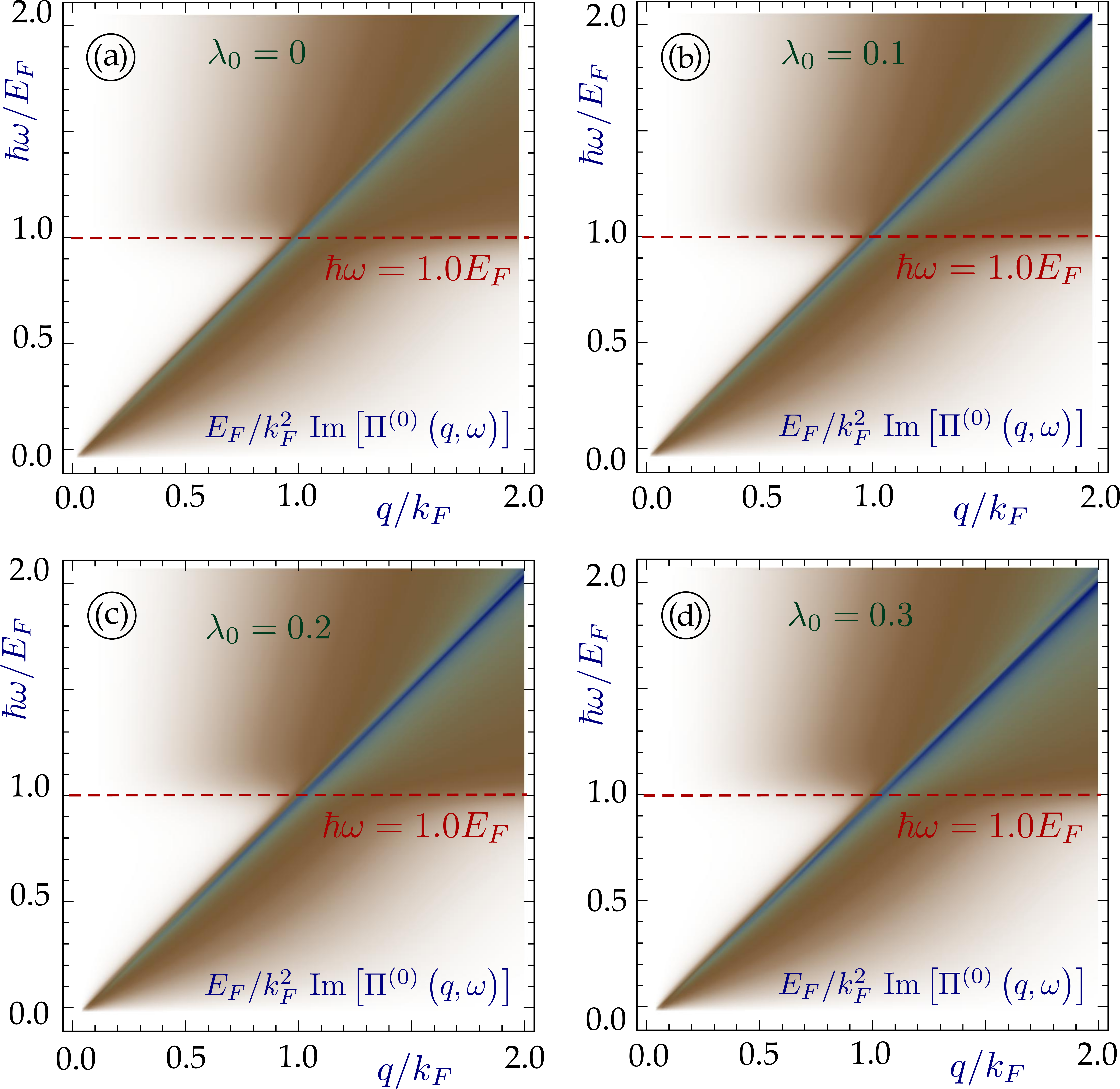}
\caption{(Color online) Density plots of $\text{Im}\left[ \Pi^{(0)}(q, \omega \, 
\vert \, \lambda_0) \right] \neq 0$ in the units of $k_F^2/E_F$ for an irradiated dice lattice with $\phi = \pi/4$. 
Each panel corresponds to a different light-electron coupling strength $\lambda_0$ of the imposed irradiation: 
$\lambda_0 = 0$ in $(a)$, $\lambda_0 = 0.1$ in $(b)$, $\lambda_0 = 0.2$ in $(c)$, and $\lambda_0 = 0.3$ in $(d)$.}
\label{FIG:5}
\end{figure}
\begin{figure} 
\centering
\includegraphics[width=0.5\textwidth]{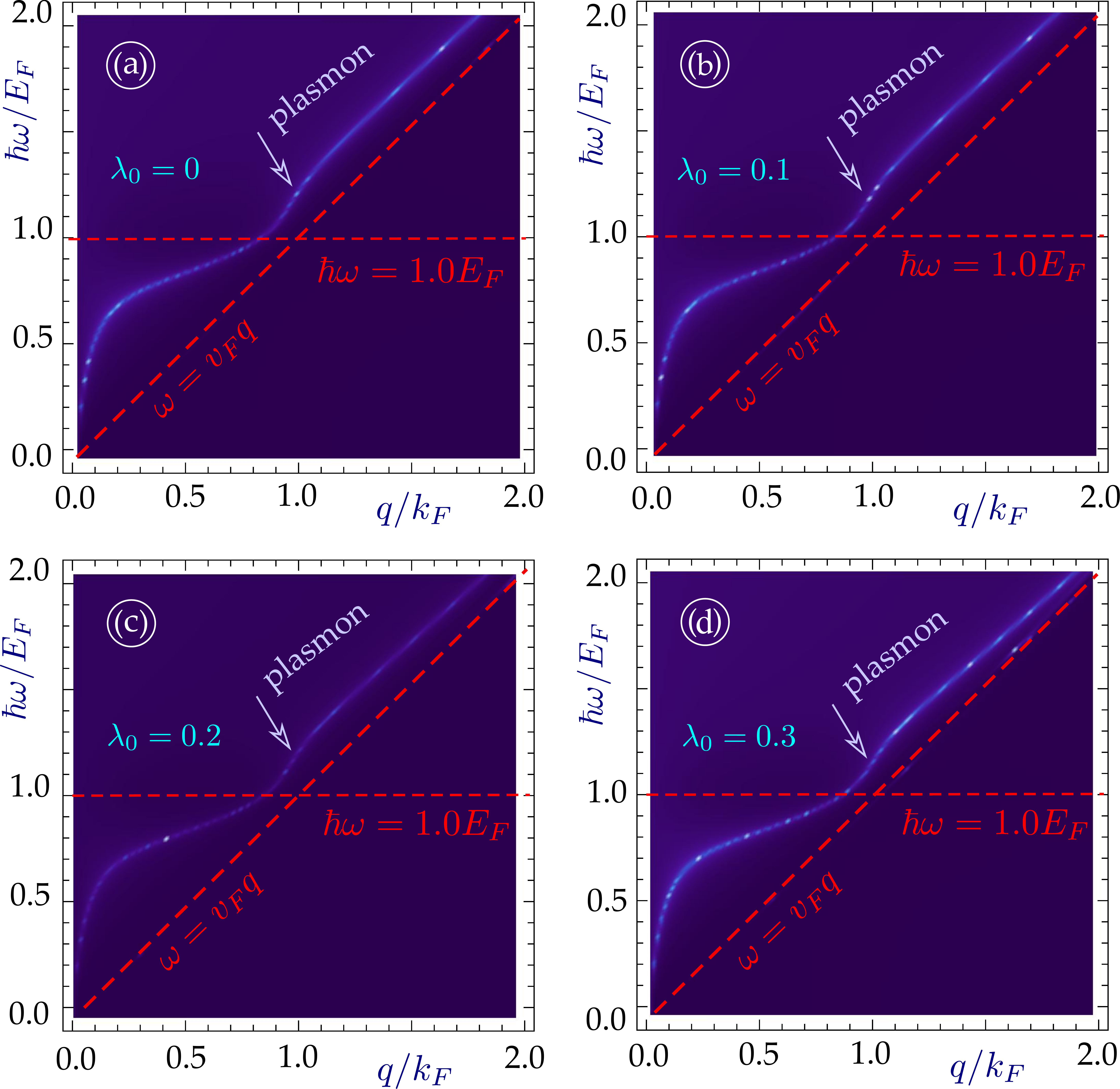}
\caption{(Color online) Density plots of $\Pi_{RPA}(q, \omega \, \vert \, \lambda_0)$ in the units of $k_F^2/E_F$ 
for an irradiated dice lattice with $\phi = \pi/4$. Here, the plasmon dispersion (blue curve) is mapped out by the peak in  correspond to the peak in 
$\left|\Pi_{RPA}(q, \omega \, \vert \, \lambda_0)\right|$ as a function of $\omega$ for each given $q$.
Four panels correspond to different electron-light coupling strengths $\lambda_0$ of the imposed irradiation:  
$\lambda_0 = 0$ in $(a)$, $\lambda_0 = 0.1$ in $(b)$, $\lambda_0 = 0.2$ in $(c)$, and $\lambda_0= 0.3$ in $(d)$.}
\label{FIG:6}
\end{figure}
\begin{figure} 
\centering
\includegraphics[width=0.5\textwidth]{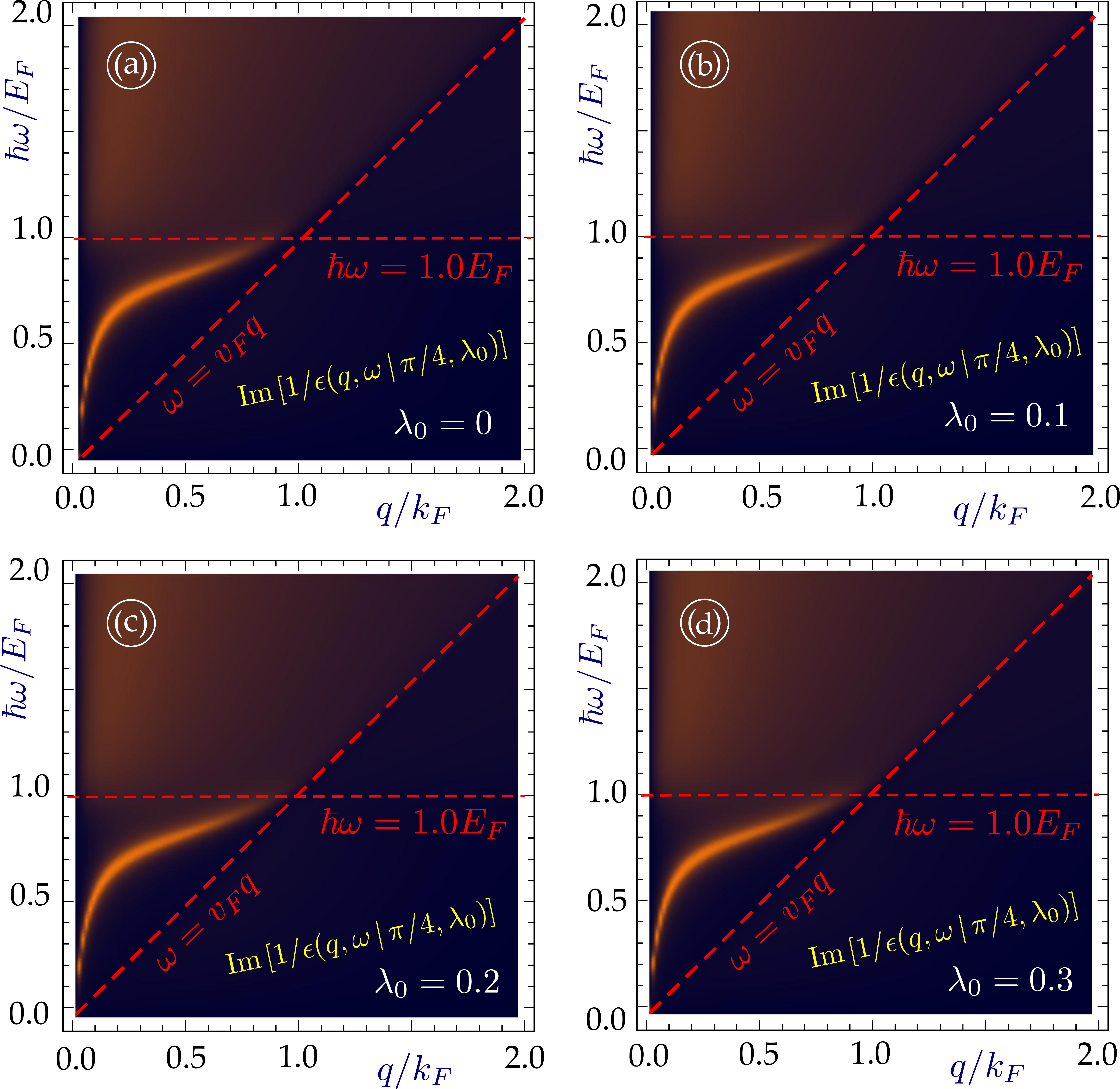}
\caption{(Color online) Density plots of energy loss function $\text{Im}\left[ 1/\epsilon(q, \omega \, \vert \, \lambda_0) \right]$ 
for an irradiated dice lattice with $\phi = \pi/4$. Here, each panel corresponds to a different electron-light coupling strength $\lambda_0$: 
$\lambda_0=0$ in $(a)$, $\lambda_0 = 0.1$ in $(b)$, $\lambda_0 = 0.2$ in $(c)$, and $\lambda_0 = 0.3$ in $(d)$.}
\label{FIG:7}
\end{figure}

Next, we turn to discussing the collective properties of an irradiated dice lattice with $\alpha = 1$. Figures \ref{FIG:5}, \ref{FIG:6}
and \ref{FIG:7} display respectively, the numerical results for particle-hole modes, plasmon branches and energy loss functions for a dice lattice irradiated by circularly-polarized light 
with different light intensities $c_0$ and light-electron coupling strengths $\lambda_0$, which provide a direct comparison with Figs.\,\ref{FIG:2}, \ref{FIG:3} and \ref{FIG:4} 
in the absence of a dressing field, i.e., taking $\lambda_0=0$.
\medskip

From Fig.\,\ref{FIG:5} we find that $\text{Im}\left[\Pi^{(0)}(q, \omega \, \vert\, \lambda_0)\right]$ for $\alpha=1$ becomes zero only below the Fermi level and above the main diagonal $\omega=v_Fq$. 
The boundary of particle-hole modes around the Fermi energy $\omega =E_F/\hbar$ does not acquire a noticeable dependence on $\lambda_0$. 
The only visible difference due to dressing field is the ``blue line'', or the negative peak in $\text{Im}\left[ \Pi^{(0)}(q, \omega \, \vert\, \lambda_0) \right]$, shifts towards under the main diagonal,  
as found from Fig.\,\ref{FIG:5}$(d)$. 
This behavior is related to light-induced modification to the energy dispersion in Eq.\,\eqref{zero} which occurs in the denominator of Eq.\,\eqref{pi0at3}. 
Physically, the change of a dressed-state energy arises from both light renormalization of the Fermi velocity and creation of an energy bandgap for nonzero $\alpha$. 
Importantly, these two modifications are of the same order of magnitude $\backsim \lambda_0$, which is quite different from graphene with $\alpha=0$ 
where only the light-induced bandgap plays a role.\,\cite{kiMain}  
The pushing-down of the main diagonal in Fig.\,\ref{FIG:5}$(d)$ can be attributed to the light renormalization of Fermi velocity in $\alpha$-$\mc{T}_3$ lattice.
This unique feature has a huge influence on suppressing the Landau damping of plasmons by decaying into particle-hole modes as shown in Fig.\,\ref{FIG:6}$(d)$, 
where the brightness of blue curves (representing strength of plasmon modes) in Fig.\,\ref{FIG:6} increases slightly with $\lambda_0$ in $(a)$ and $(b)$, followed by a dramatic decrease in $(c)$, 
and ended with a huge increase again in $(d)$.
\medskip

The feature observed in Fig.\,\ref{FIG:5} is also reflected in the energy loss function $\mc{L}(q,\omega \, \vert \, \lambda_0)$ given by Eq.\,\eqref{eloss} for $\alpha=1$. 
While the undapmed plasmon shows a small decrease of its frequency for $q/k_F<1$ and $\hbar \omega/E_F<1$ as seen in $(a)$ - $(c)$ of Fig.\,\ref{FIG:7}, 
the energy-loss curve or the undamped plasmon branch approaches the main diagonal for a larger $\lambda_0=0.3$ in $(d)$. 
On the other hand, the damped plasmon branch for $q/k_F>1$ tends to follow the blue line in Fig.\,\ref{FIG:5} or the lines of the poles of both real and imaginary parts of 
$\Pi^{(0)}(q, \omega \, \vert\, \lambda_0)$ in Eq.\,\eqref{pi0at3}. 
As $\lambda_0$ increases, both the blue line and plasmon branch reduces their frequencies and becomes located below the main diagonal $\omega = v_F q$. 
The change of $\Pi^{(0)}(q, \omega \, \vert\, \lambda_0)$ and plasmon branch under the irradiation is quite conspicuous even though we limit our consideration to the small values 
of $\lambda_0 \ll 1$ for the off-resonant dressing field. 
The ``kinky'' shape of the plasmon branch is preserved and remains nearly the same in despite of the external irradiation.       	
	
\section{Optical Conductivity in $\alpha$-$\mathcal{T}_3$ Lattice}
\label{sec5}

Optical conductivity connects the electric properties of an low-dimensional material, such as polarization current,
with incident optical field. Therefore, we expect that
optical conductivity will closely related to other optical properties of materials, such as absorption.  
Explicitly, the optical-current conductivity $\sigma_O(\omega\,\vert\, \phi,\lambda_0)$ can be calculated from the polarization function $\Pi^{(0)}(q,\omega \,\vert \, \phi,\lambda_0)$
in Eq.\,\eqref{pi0at3} under the long-wavelength limit,\,\cite{SilMain} yielding

\begin{equation}
\label{sigmaO}
\sigma_O(\omega\,\vert\, \phi,\lambda_0) = i e^2\omega \, \lim\limits_{q \to 0}\left[\frac{1}{q^2} \,\, \Pi^{(0)}(q,\omega \,\vert \, \phi,\lambda_0)\right]\ . 
\end{equation}
From Eq.\,\eqref{sigmaO}, it is clear that the real part of $\sigma_O(\omega\,\vert\, \phi,\lambda_0)$ is associated with the imaginary part of $\Pi^{(0)}(q,\omega \,\vert \, \phi,\lambda_0)$ 
for single-particle excitation, while the imaginary part of $\sigma_O(\omega\,\vert\, \phi,\lambda_0)$ corresponds to the real part of $\Pi^{(0)}(q,\omega \,\vert \, \phi,\lambda_0)$ 
for induced optical polarization. 
\medskip

\begin{figure} 
\centering
\includegraphics[width=0.6\textwidth]{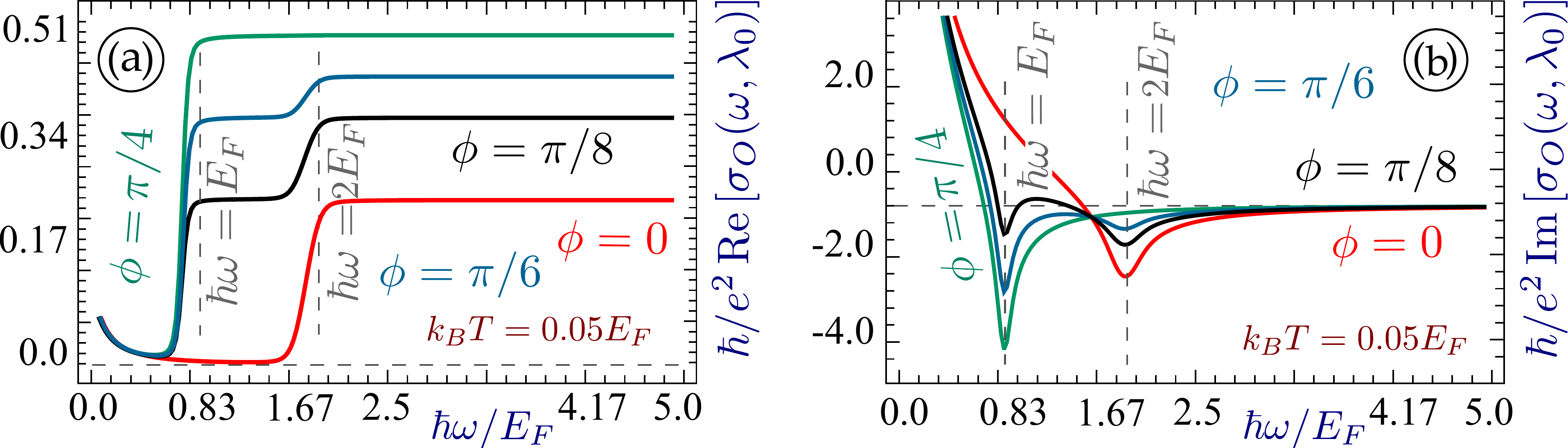}
\caption{(Color online) Optical conductivity $\sigma_O(\omega\,\vert\, \phi,\lambda_0)$ in the units of $e^2/\hbar$ as a function of
frequency $\omega$ at low temperature $k_BT/E_F=0.05$. Panels $(a)$ and $(b)$ show the real and imaginary parts of $\sigma_O(\omega\,\vert\, \phi,\lambda_0)$, respectively.
In both panels, each colored curve relates to a specific value of $\phi=\tan^{-1}\alpha$: red for $\phi = 0$ (graphene), black for $\phi = \pi/8$, 
blue for $\phi = \pi/6$, and green for $\phi = \pi/4$ (dice lattice).}
\label{FIG:8}
\end{figure}

The calculated real part of $\sigma_O(\omega\,\vert\, \phi,\lambda_0)$ is presented Fig.\,\ref{FIG:8}$(a)$, which describe the absorptive dissipation of polarization current or 
the damping of plasmon mode determined by the imaginary part of $\Pi^{(0)}(q,\omega \,\vert \, \phi,\lambda_0)$ in Eq.\,\eqref{pi0at3}.
From Fig.\,\ref{FIG:8}$(a)$ we find the leading threshold step for the damping of plasmon excitation at $\omega=E_F/\hbar$ as $\alpha\neq 0$. 
For $\alpha=0$, however, this leading threshold step shifts up to $\omega=2E_F/\hbar$ for graphene.
Additionally, numerical results for the imaginary part of $\sigma_O(\omega\,\vert\, \phi,\lambda_0)$ is displayed in Fig.\,\ref{FIG:8}$(b)$, 
which leads to the dissipation of optical current due to the induced polarization field given by the real part of $\Pi^{(0)}(q,\omega \,\vert \, \phi,\lambda_0)$ 
or the plasmon excitation in our considered system. 
In a correspondence to the threshold steps for plasmon damping in Fig.\,\ref{FIG:8}$(a)$, from Fig.\,\ref{FIG:8}$(b)$ we find negative peaks at these frequencies 
related to starting of the undamped plasmon excitations.
Interestingly, we also observe from Fig.\,\ref{FIG:8}$(b)$ the secondary higher-frequency graphene-like negative peak even for $\phi\neq \pi/4$, which is associated with 
the appearance of the secondary higher-frequency threshold step in Fig.\,\ref{FIG:8}$(a)$. 
Since the real and imaginary parts of $\Pi^{(0)}(q,\omega \,\vert \, \phi,\lambda_0)$ in Eq.\,\eqref{pi0at3} are related to each other by Krammers-Kronig relations, 
we should expect to see corresponding dependence between the two parts of $\sigma_O(\omega\,\vert\, \phi,\lambda_0)$, as shown in Figs.\,\ref{FIG:8}$(a)$ and \ref{FIG:8}$(b)$. 
Here, the strong negative peaks in $(b)$ for $\text{Im}\left[\sigma_O(\omega\,\vert\, \phi,\lambda_0) \right]$ and the abrupt sudden jumps in $(a)$ for $\text{Im}\left[\sigma_O(\omega\,\vert\, \phi,\lambda_0) \right]$ occur exactly at the same frequencies.  
\medskip 	

Physically, as seen from Fig.\,\ref{FIG:8}$(a)$, we would like to point that 
the behavior for zero real part of $\sigma_O(\omega\,\vert\, \phi,\lambda_0)$ as a function of $\omega$ below a certain value is termed as 
Pauli-blocking effect.\,\cite{addG1, ourCond18PRB} This threshold frequency depends primarily on the structural parameter $\alpha$ for different $\alpha$-$\mc{T}_3$ lattices 
and originates from its limiting behavior of the imaginary part of $\Pi^{(0)}(q,\omega \,\vert \, \phi,\lambda_0)$ in Eq.\,\eqref{pi0at3} as $q\to 0$. 
For $0 < \alpha < 1$, this exists a two-step process, i.e., two sudden jumps in $\text{Re} \,\left[\sigma_O(\omega\,\vert\, \phi,\lambda_0) \right]$ at $\hbar\omega/E_F=1,\,2$. 
The first jump arises from the $ 0 \Longleftrightarrow 1$ transitions of electrons from (and to) the flat band and can be represented by $\gamma = 0$ and $\gamma' = 1$ (or $\gamma = 1$ and $\gamma' = 0$) 
terms in their summations in Eq.\,\eqref{pi0at3}. 
These transitions exist only in the presence of the flat band and disappear for graphene with $\alpha=0$, in which only one jump occurs in $\text{Re}\left[\sigma_O(\omega\,\vert\, \phi,\lambda_0)\right]$ 
from the $-1 \Longleftrightarrow +1$ transitions, or $\gamma = \pm 1$ and $\gamma' =\mp 1$, part of their summations in Eq.\,\eqref{pi0at3}. 
Similarly, the boundary for particle-hole mode is found sitting at $\omega = 2 E_F/\hbar - v_Fq$ for graphene but at $E_F/\hbar - v_Fq$ for all other non-zero $\alpha$ values.         
\medskip

\section{Coupled Plasmons in $\alpha$-$\mc{T}_3$ Double Layers}
\label{sec6}

In this section, we investigate the properties of the dispersion relations of coupled plasmon modes in an interacting double layer $\alpha$-$\mc{T}_3$ lattices in the absence of light irradiation.  In this case, we expect that the individual plasmon modes in each of the two layers will be modified by the interlayer Coulomb interaction, and the whole system behaves in a way similar to two coupled quantum oscillators. 

\medskip

Plasmon excitations in two spatially separated and interacting layers with a two dimensional electron gas were investigated initially and reported in Ref.\,\cite{opg}, and the coupled plasmon modes in an arbitrary number $N$ of graphene layers interacting with a semi-infinite conductor were analyzed in Refs.\,\cite{ourJAP2017} and \cite{ourSREP}. Generally, the total number of coupled-plasmon modes in such a system is equal to the number of conducting layers, i.e., $N+1$ if the surface-plasmon mode in the semi-infinite conductor is also taken into consideration.  For this case, each plasmon branch is often partially Landau damped at some critical frequency and wave vector.  If only $N$ two dimensional layers without a conducting surface are considered, the plasmon dispersions will be determined by a linear and homogeneous $(N \times N)$ matrix equation, where each individual equation is coupled to all other ones by the inter-layer potential as given by Eq.\,\eqref{Co12}.

\medskip

Specifically, for a double layer system composed of two Coulomb interacting $\alpha$-$\mc{T}_3$ lattices separated by a distance $d$ from each other, the resulting determinant equation takes the form

\begin{eqnarray}
\nonumber
&&  \left\{
1 - \mbb{V}_0(q) \, \Pi^{(0)}(q, \omega \, \vert \, \phi_1)
\right\} \,
\left\{
1 - \mbb{V}_0(q) \, \Pi^{(0)}(q, \omega \, \vert \, \phi_2)
\right\} \\
\label{Madh}
&& -  \left[ \mbb{V}_1(q)\right]^2 \, \Pi^{(0)}(q, \omega \, \vert \, \phi_1) \, \Pi^{(0)}(q, \omega \, \vert \, \phi_2) = 0 \ ,  
\end{eqnarray}
where $\Pi^{(0)}(q, \omega \, \vert \, \phi_{1,2})$ represent the polarization functions of two layers labeled by $i=1,2$, and  the intra-layer (0) and inter-layer (1) Coulomb potentials are found as

\begin{eqnarray}
\nonumber
&& \mbb{V}_0(q) = \frac{e^2}{2\epsilon_0\epsilon_rq} \, , \\
\label{Co12}
&& \mbb{V}_1(q) = \frac{e^2}{2\epsilon_0\epsilon_rq} \, \tet{exp}(- q d) \ .
\end{eqnarray}
Here, we assume that the Fermi energies in both layers is equal to each other, i.e., $E_{F,\,1} = E_{F,\,2}$, and the two layers are otherwise identical except for having different phases $\phi_1 \neq \phi_2$ (or $\alpha_1 \neq \alpha_2$). 

\medskip

\begin{figure}
\centering
\includegraphics[width=0.8\textwidth]{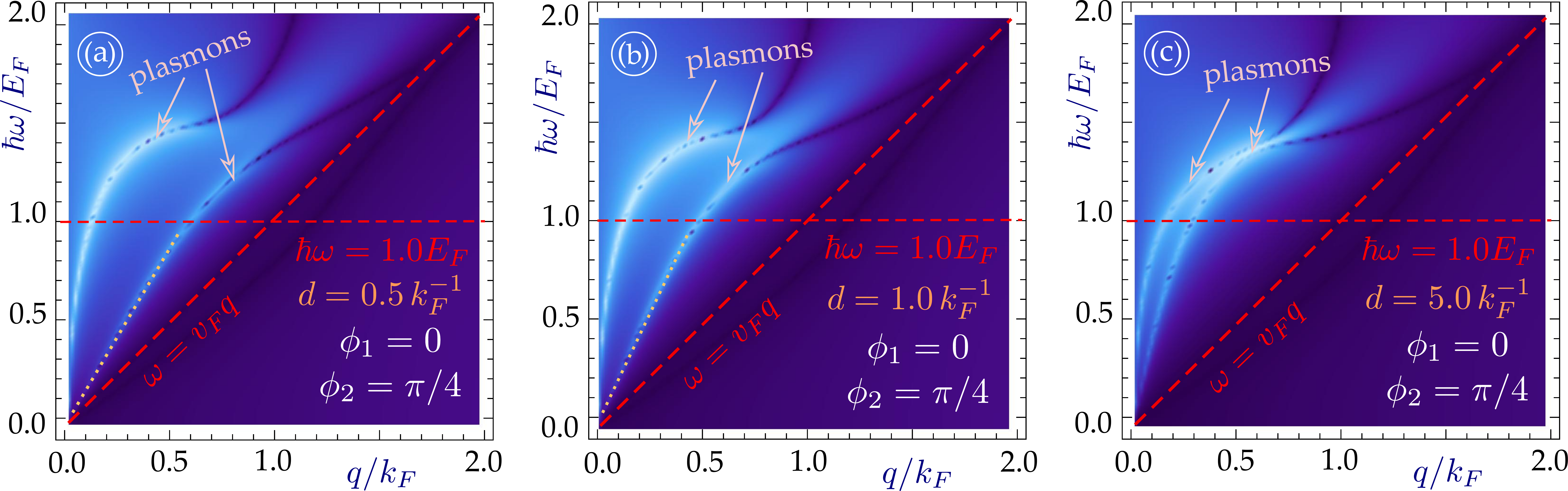}
\caption{(Color online) Density plots of plasmon dispersion relations calculated from Eq.\,\eqref{Madh}
for two Coulomb-coupled dice lattice layers with $\phi_1 =0$ and $\phi_2 = \pi/4$. Here, each panel corresponds to chosen layer separation: $d = 0.5\,k_F^{-1}$ in $(a)$, $d=1.0\,k_F^{-1}$ in $(b)$, and $d=5.0\,k_F^{-1}$ in $(c)$.}
\label{FIG:9}
\end{figure}

\begin{figure}
\centering
\includegraphics[width=0.8\textwidth]{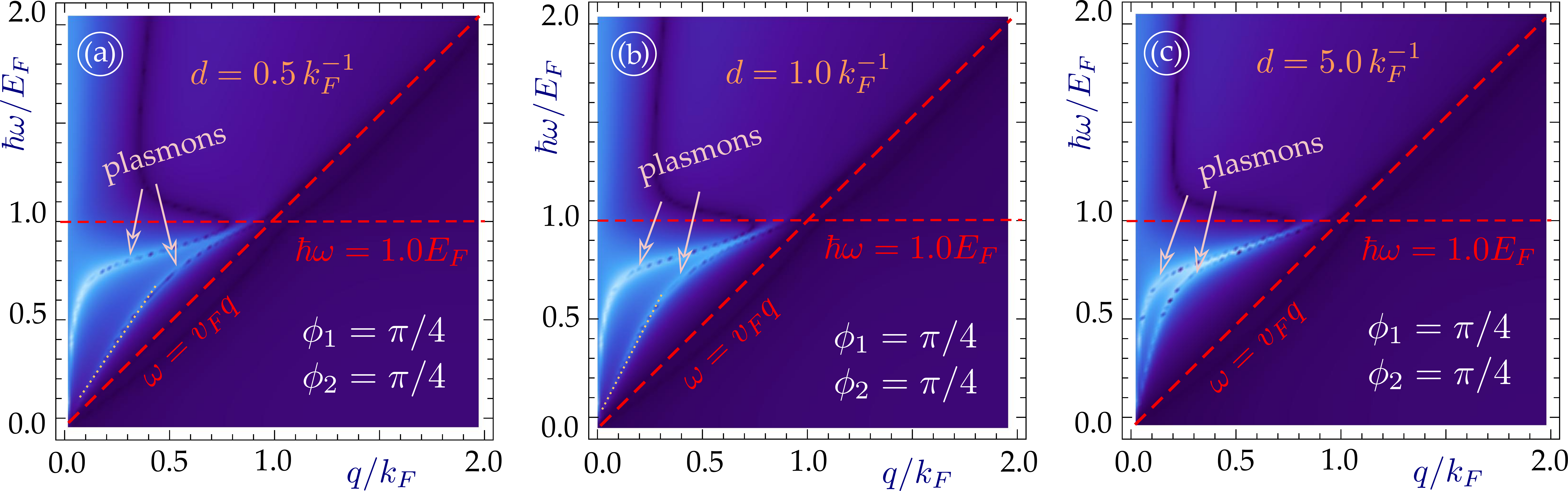}
\caption{(Color online) Density plots of plasmon dispersions based on Eq.\,\eqref{Madh}
for two Coulomb-coupled layers composed of dice lattices with $\phi_1 =\phi_2 = \pi/4$. Here, each panel corresponds to the following separations between the two layers: $d = 0.5\,k_F^{-1}$ in $(a)$, $d=1.0\,k_F^{-1}$ in $(b)$, and $d=5.0\,k_F^{-1}$ in $(c)$.}
\label{FIG:10}
\end{figure}

Our numerical results for coupled plasmon modes within two interacting layers of $\alpha$-$\mc{T}_3$ lattices are presented in Figs.\,\ref{FIG:9} and \ref{FIG:10}.  Two different layers are chosen in Fig.\,\ref{FIG:9} with their phases $\phi_1 =0$ and $\phi_2 = \pi/4$ for graphene and dice lattice, respectively,  whereas the same dice lattice is chosen in both layers for Fig.\,\ref{FIG:10}. The presence of the inter-layer potential $\mbb{V}_1(q)$ substantially modifies the dispersions of these two coupled  plasmon modes with  a splitting gap as the layer separation $d$ is decreased, as it is evident in Fig.\,\ref{FIG:9}$(a)$.  In fact, the interlayer interaction plays a role only for separations up to $d \approx k_F^{-1}$. However, it is rapidly decreased once the layer separation is increased and reaches $d = 5.0\,k_F^{-1}$.  Then, the off-diagonal terms in Eq.\,\eqref{Madh} become negligibly small, as demonstrated in both Figs.\,\ref{FIG:9}$(c)$ and \ref{FIG:10}$(c)$.

\medskip

The most crucial finding from Figs.\,\ref{FIG:9} and \ref{FIG:10} is that {\it the lower acoustic plasmon branch remains linear\/} despite the choice of the phase $\phi_i$ or separation $d$ between two layers as long as the second term in Eq.\,\eqref{Madh} remains substantial.  This result is similar to that found for graphene reported  in Ref.\,\cite{opg}. On the contrary, the higher branch always displays a $\backsim \sqrt{q}$ dependence and therefore reaches the regions for particle-hole excitations at smaller $q$ values.

\medskip

The plasmon damping is determined by the outermost boundary of the particle-hole modes which corresponds to a larger phase $\phi$. Therefore, all plasmon branches are free from damping inside the triangle region determined by $\omega < E_F/\hbar$ and $\omega > v_F q$, as we discussed above.  Additionally,  when we compare Fig.\,\ref{FIG:9} with Fig.\,\ref{FIG:10}, we find much stronger  damping for the plasmon excitations  above the line $\omega=E_F/\hbar$ in Fig.\,\ref{FIG:10} as two layers are assumed to be the same with $\phi_1=\phi_2$, and two plasmon branches merging together as $q$ gets close to $k_F$.Technically, we can always achieve a lower slope for the linear plasmon branch by varying the  separation $d$ between the two layers.  At the same time,a more pronounced change in the plasmon dispersion could also be achieved by varying the Fermi energy $E_{F,i}$ for one of the two layers,  since the plasmon frequency is nearly proportional to the Fermi energy in the long-wavelength limit.

\section{Optical Absorption in $\alpha$-$\mc{T}_3$ Multi-Layer System}
\label{sec7}

The degree of optical absorption, or the loss of photons, in a system can be quantified by the so-called dimensionless absorbance $\Gamma_{abs}(\omega)$,\,\cite{DanR41, ggb} which is calculated as

\begin{equation}
\label{betaAbs}
\Gamma_{abs}(\omega) = \frac{\omega\sqrt{\epsilon_r}}{c} \, \left[1 + \rho_{ph}(\omega) \right] \,\text{Im}\left[
\alpha (\omega)
\right] \ ,
\end{equation}
where $\hbar\omega$ is the energy of incident photons. At low temperatures, i.e., $k_BT\ll \hbar\omega$, the photon occupation number $\rho_{ph}(\omega)$  becomes negligible due to

\begin{equation}
\rho_{ph}(\omega) = \left[
\tet{exp} \left(\frac{\hbar \omega}{k_B T} \right) - 1
\right]^{-1} \backsim \tet{exp} \left[ - \frac{\hbar \omega}{k_B T} \right] \ll 1 \ .
\end{equation}
On the other hand, the Lorentz like factor (in units of length) introduced in Eq.\,\eqref{betaAbs} is calculated as\,\cite{DanR12}

\begin{equation}
\label{alphaL1}
\alpha(\omega) = - \left(  \frac{2 e^2}{\epsilon_0 \epsilon_r} \right) \, \pi R_0^2 \, \sum\limits_{j=1}^{N} \,
\int \frac{d^2\mbox{\boldmath$q$}}{(2 \pi)^2} \, \tet{exp} \left\{ - \frac{q^2R_0^2}{4}  \right\} \, \mc{Q}_j(q,\omega\,\vert\,\{\phi\}) \ ,
\end{equation}
where the transverse linearly polarized light field is incident perpendicularly to layers with a beam radius $R_0$. The optical response function $\mc{Q}_j(q,\omega\,\vert\,\{\phi\})$ in Eq.\,\eqref{alphaL1} for the $j$'s layer in an $N-$layer system has been modified for the  Dirac cone dispersion from its original expression given in Ref.\,\,\cite{DanR42}, leading to

\begin{equation}
\mc{Q}_j(q,\omega\,\vert\,\{\phi\}) = \chi_{j,j}(q, \omega\,\vert\,\{\phi\}) \, \left( \frac{v_F q}{k_F \omega} \right)^2  \ ,
\label{Qpar}
\end{equation}
where $\{\phi\}$ represents the full set $\{\phi_1,\,\phi_2,\,\ldots,\,\phi_{N-1},\,\phi_N\}$, $v_F$ and $k_F=\sqrt{\pi n_0}$ are Fermi velocity and wave number of Dirac electrons, and only electron doping in the conduction band is assumed with an areal density $n_0$.  Moreover, $\chi_{j,j'}(q, \omega\,\vert\,\{\phi\})$ introduced in Eq.\,\eqref{Qpar} can be calculated from the regular non-interacting  polarization function $\Pi_j^{(0)}(q,\omega \, \vert \, \phi_j)$  in Eq.\,\eqref{pi0at3} for the $j$'s layer by taking into account the Coulomb coupling between various $\alpha$-$\mc{T}_3$ layers. This yields

 \begin{eqnarray}
\nonumber
\chi_{j,j'}(q, \omega\,\vert\,\{\phi\}) &=& \Pi_j^{(0)}(q, \omega\,\vert\,\phi_j)\,\delta_{j,j'}\\
& +& \Pi^{(0)}_j(q, \omega\,\vert\,\phi_j) \sum\limits_{s (\neq j') = 1}^{N}
\mbb{V}_{j,s}(q)\, \chi_{s,j'}(q, \omega\,\vert\,\{\phi\}) \ ,
\label{chi0}
\end{eqnarray}
where $\mbb{V}_{j,s}(q)=(e^2/2\epsilon_0\epsilon_rq)\,\exp(-qd|s-j|)$ represents either interlayer ($j\neq s$) or intralayer ($j=s$) Coulomb interaction. Equation\ \eqref{chi0} is physically equivalent to the random-phase approximation for including the screening from both intra-layer and inter-layer Coulomb interactions in a multi-layer system. By solving Eq.\,\eqref{chi0} to determine $\chi_{j,j}(q, \omega\,\vert\,\{\phi\})$, and then employing Eqs.\,\eqref{alphaL1} and \eqref{Qpar}, we eventually arrive at a result for the absorbance $\Gamma_{abs}(\omega)$  in Eq.\,\eqref{betaAbs}.

\section{Concluding remarks and summary}
\label{sec8}

We have investigated fundamental characteristics of the collective, transport and optical properties of electron dressed states  in single  and double layer $\alpha$-$\mc{T}_3$ materials. These dressed electronic states, due to the interaction between electrons and  incident  circularly polarized light field within the off-resonant regime, acquire new light renormalized energy dispersion.   Additionally, this leads to modified Fermi velocity and opening an energy gap much smaller than that for irradiated graphene. It is also worthy of note that  we have carefully analyzed the role played by many-body effects on plasmon excitations, and the way in which they are affected  by the hopping parameter $\alpha$ as well as the electron-light coupling strength  $\lambda_0$ for a dice lattice  when $\alpha = 1$. The interplay between all these effects varies in scope and even compete with each other,  resulting in a practical opportunity for light being used for tuning the electronic and collective properties of optical materials by means of Floquet engineering.

\medskip
 
From the results of our calculations, we have discovered that all the unique properties of $\alpha$-$\mc{T}_3$ originate from the presence of a flat band in the energy dispersions and the resulting inter-band transitions. Specifically, we have focused our attention on the optical conductivity and demonstrated how the transitions between the flat and conduction bands  give rise to a step-like increase in the optical conductivity at the Fermi energy not present in graphene.

\medskip

We note our calculations of the coupled plasmon excitations for interacting double layer $\alpha$-$\mc{T}_3$ lattices.   Their dispersions reveal that the lower plasmon branch remains linear (an acoustic mode) for all $\alpha$. However, this group velocity is changed by varying the layer separation and doping level of individual layers.  Such a novel feature provides us with an undamped plasmon excitation below the Fermi level for large wave vectors.
\medskip

Finally, we have established a theoretical framework for computing the optical absorbance of doped multi-layered $\alpha$-$\mc{T}_3$ in the presence of linearly polarized light by including both inter-layer and intra-layer Coulomb interactions between Dirac electrons. All explored optical properties of these novel pseudospin-$1$ materials in this paper represent important and surprising discoveries, which could be applied in the construction of  new optical and plasmonic nanoscale devices.

\section*{Acknowledgement(s)}
G.G. would like to acknowledge the financial support from the Air Force Research Laboratory (AFRL) through 
grant FA9453-18-1-0100 and award FA2386-18-1-0120. D.H. thanks the supports from the Laboratory University
Collaboration Initiative (LUCI) program and from the Air Force Office of Scientific Research (AFOSR).

\section*{Disclosure statement}
No potential conflict of interest was reported by the authors.

\end{document}